\documentclass[lettersize,journal]{IEEEtran}
\usepackage{cite}
\usepackage{amsmath,amssymb,amsfonts}
\usepackage{algorithm,algorithmic}
\usepackage{graphicx}
\usepackage{textcomp}
\usepackage{subfigure}
\usepackage{hyperref}
\usepackage{float}
\hypersetup{hidelinks=true}
\usepackage{textcomp}
\usepackage{xcolor}
\usepackage{soul}
\usepackage{booktabs}
\usepackage[section]{placeins}

\def\BibTeX{{\rm B\kern-.05em{\sc i\kern-.025em b}\kern-.08em
    T\kern-.1667em\lower.7ex\hbox{E}\kern-.125emX}}

\begin{document}

\title{Selection and Stability of Functional Connectivity Features for Classification of Brain Disorders}

\author{Aniruddha Saha, Soujanya Hazra, and Sanjay Ghosh, \IEEEmembership{Senior Member, IEEE}
\thanks{This research is supported by the Faculty Start-up Research Grant (FSRG), IIT Kharagpur, awarded to Dr. Sanjay Ghosh.}
\thanks{Aniruddha Saha, Soujanya Hazra, and Sanjay Ghosh are with the Department of Electrical Engineering, Indian Institute of Technology Kharagpur, WB 721302, India (e-mail: \texttt{sanjay.ghosh@ee.iitkgp.ac.in}).}
}

\maketitle

\begin{abstract}
Brain disorders are an umbrella term for a group of neurological and psychiatric conditions that have a major effect on thinking, feeling, and acting.
These conditions encompass a wide range of conditions.  The illnesses in question pose significant difficulties not only for individuals, but also for healthcare systems all across the world.  In this study, we explore the capability of explainable machine learning for 
classification of people who suffer from brain disorders.
This is accomplished by the utilization of brain connection map, also referred as \textit{connectome}, derived from functional magnetic resonance imaging (fMRI) data.  In order to analyze features that are based on the connectome, we investigated several different feature selection procedures. These strategies included the Least Absolute Shrinkage and Selection Operator (LASSO), Relief, and Analysis of Variance (ANOVA), in addition to a logistic regression (LR) classifier.  First and foremost, the purpose was to evaluate and contrast the classification accuracy of different feature selection methods in terms of distinguishing healthy controls from sick or injured individuals.  The evaluation of the stability of the traits that were chosen was the second objective.  The identification of the regions of the brain that have an effect on the classification was the third main objective.  When applied to the UCLA dataset, the LASSO approach, which is our most effective strategy, produced a classification accuracy of 91.85\% and a stability index of 0.74, which is greater than the results obtained by other approaches: Relief and ANOVA.  These methods are effective in locating trustworthy biomarkers, which adds to the development of connectome-based classification in the context of issues that impact the brain. 
\end{abstract}

\begin{IEEEkeywords}
Biomarker, brain disorders, feature stability, functional connectivity, logistic regression classifier.
\end{IEEEkeywords}

\section{Introduction}
\label{sec:introduction}
\IEEEPARstart{B}{rain} disorders represent a diverse group of neurological and psychiatric conditions that severely impact an individual’s cognitive functions, emotional regulation, and perception of reality \cite{ma2023multi}. Characterized by symptoms such as cognitive impairments, mood disturbances, abnormal behavior, and disruptions in sensory processing, these conditions pose complex challenges in both diagnosis and treatment \cite{WHITEFORD20131575}. The exact etiologies of many brain disorders remain elusive, although it is widely acknowledged that a multifactorial interplay of genetic, environmental, and neurobiological factors contributes to their onset \cite{Zhang2023}. Brain connectivity map or connectome is distinct in several ways, presenting both unique challenges and opportunities when applied in machine learning. Unlike conventional medical images, which have a grid-like structure where each pixel is only adjacent to nearby pixels, connectomes possess a more complex topology. In connectome data, each brain region, represented as a node in the network, can be connected to any other region, either structurally or functionally \cite{Lynall9477}. It is a matrix-like structure where the size of the matrix, i.e, rows or columns, represents the number of brain regions or Region of Interest (ROI) under study. These connections are represented by binary or weighted edges between nodes. This network structure encapsulates essential information about brain connectivity that is not directly captured by other imaging modalities. 

Structural magnetic resonance imaging (sMRI) is a popular tool for studying brain alterations, since it offers comprehensive anatomical images of the brain \cite{COCCHI2014779}. It makes it possible to build structural connectomes, which record the brain's physical structure. Functional MRI (fMRI) measures blood-oxygen-level-dependent signals, which indicate regional blood flow changes as proxies for neuronal co-activation \cite{Lynall9477}. The synchronized activations across spatially different brain areas constitute the foundation of functional connectivity (FC). It is a network-level representation commonly employed to investigate illnesses including Alzheimer’s disease \cite{hao2022multimodal}, major depressive disorder \cite{hasanzadeh2020graph}, autism \cite{barik2023functional}, and schizophrenia \cite{Lynall9477}.
Although fMRI is the predominant method for FC analysis, alternative techniques like electroencephalography (EEG) \cite{hasanzadeh2020graph} and magnetoencephalography (MEG) \cite{Jin2024} provide supplementary insights with superior temporal resolution. FC data is highly dimensional and complicated, so successful classification models require great care in feature selection and stability. It needs to make sure the biomarkers found are reliable and have an impact on the nervous system.

In this paper, we investigate three different feature
selection procedures to classify brain disorders based on functional connectivity. We also analyze the stability of these selected features in terms of how consistently the same features are selected across different iterations, datasets, or
models.
Our research expands on previous efforts to develop stable biomarkers in brain connectomes \cite{GUTIERREZGOMEZ2020102316} to assess the classification of brain disease based on functional connectivity. Unlike classification-only approaches, we prioritize feature stability, interpretability, and reproducible findings. In particular, the key contributions of this paper are:
\begin{itemize}
    \item We evaluate and compare three popular feature selection algorithms, LASSO, Relief, and ANOVA, on two fMRI datasets, analyzing a total of 16,769 functional connectivity features per subject (3,403 in Dataset 1 and 13,366 in Dataset 2).
    
    \item We specifically quantify feature stability across cross-validation folds using the \textit{Kuncheva index} and \textit{Jaccard index} to ensure biomarker discovery reproducibility. The average stability was achieved by LASSO (\textit{Kuncheva}: 0.74 and \textit{Jaccard}: 0.69).
    
    
    \item We illustrate the generalizability of our findings on two different fMRI datasets with a total of 226 subjects (54 in Dataset 1 and 172 in Dataset 2), demonstrating consistent classification accuracy and feature robustness trends. LASSO achieved up to 91.85\% accuracy and 91.98\% F1-score.
\end{itemize}

This paper is summarized as follows. Section \ref{sec:related_work} discusses neuro-imaging functional connectivity analysis, feature selection, and stability assessment.  Section \ref{sec:method} covers the datasets, feature extraction pipeline, classification model, feature selection techniques, and stability metrics. Section \ref{sec:results} presents the experimental setup, classification findings, and feature stability analysis across datasets. Section \ref{sec:explain} analyzes explainability by identifying consistently identified brain areas that contribute to classification.  Section \ref{sec:conc} closes the paper and suggests future research.


\section{Related Work}
\label{sec:related_work}
Over the past decade, numerous studies have investigated the correlation between brain functional connectivity (FC) and various neurological and mental problems. Researchers identified, selected, and interpreted neuro-imaging characteristics from fMRI \cite{Lynall9477}, EEG \cite{hasanzadeh2020graph}, and MEG \cite{Jin2024} for diagnostic and prognostic reasons using machine learning and deep learning. These systems demonstrate that feature selection, graph-based representations, and multi-modal fusion improve brain disease accuracy and interpretability.
Qu et al. \cite{chen2021estimation} introduced a deep graphical approach to extract distinguishing features from functional brain networks. A nonlinear network fusion method using message-passing mechanisms was used to assess multimodal brain connectivity \cite{qu2021brain}. Jie et al. \cite{jie2013integration} improved categorization using topological and statistical FC features across neuroimaging tasks. According to the authors in this study \cite{kragel2022temporal}, spontaneous emotional brain states evolve and affect mental health. Tawhid and colleagues introduced GENet \cite{tawhid2024genet}, a generic neural model for EEG-based neurological diseases. Dai et al. \cite{dai2025feature} suggested a transfer learning method that aligns features across domains to enhance the classification of psychiatric disorders. Authors in \cite{qin2025classification} employed a region-selected graph convolutional network (GCN) to dynamically identify disorder-specific patterns instead of relying on fixed brain areas. A multi-scale dynamic graph technique to model FC fluctuations over time was introduced in another study \cite{ma2023multi}.
Multimodal integration and graph-based learning have been the primary areas of attention in research about the classification of Alzheimer's disease. An approach to learning that is self-paced and makes use of structural and functional imaging was proposed in \cite{hao2022multimodal}. The application of EEG-based graph neural network (GNN) was demonstrated in \cite{klepl2022eeg}, which showed that the selection of the FC metric had a considerable impact on classification performance.
Cao and collaborators \cite{cao2023novel} introduced a dynamic connectivity-based approach for more precise detection of mild cognitive impairment (MCI). In a parallel study, \cite{ma2022diagnosis}  presented an ordinal pattern kernel that captures brain signal temporal patterns for MCI classification.


Kang et al. \cite{kang2023classifying} identified depressed patients using residual neural networks trained on certain frequency bands and brain areas. Another study \cite{zheng2023attention} presented an attention-based fusion model for depression diagnosis using multimodal MRI data. The work by Hasanzadeh and colleagues \cite{hasanzadeh2020graph} utilized graph theory to study directed FC networks from EEG signals. Guo et al. \cite{guo2020diagnosis} built whole-brain effective connection networks from resting-state fMRI to find reliable depression related patterns.
The innovative schizophrenia diagnosis framework, Schizo-Net \cite{grover2023schizo}, uses multimodal deep learning and EEG-based connection biomarkers.
A study \cite{huang2022joint} provided a strategy for selecting joint channels and connectivity applied to fNIRS data for stress detection in decision-making scenarios.
The authors \cite{barik2023functional} utilized machine learning algorithms on MEG data to detect autism in children, with a focus on early diagnosis. In order to highlight the temporal dynamics linked to autism, Jamal et al. \cite{jamal2014classification} used supervised learning to analyze synchrostate-derived EEG characteristics.
The recent paper \cite{han2024brain} examined how dynamic FC changes with age could predict brain age. Their connection analysis method may detect early cognitive deterioration.


Neuroimaging investigations require feature selection due to the connectome data's high dimensionality.  Ji and Yao \cite{ji2020convolutional} suggested a CNN-graphical Lasso model for sparse, topologically significant features. The authors in \cite{GUTIERREZGOMEZ2020102316} developed a stability-driven framework to find repeatable biomarkers from functional connectomes in schizophrenia, highlighting the necessity for rigorous feature selection. In the same direction, the study in \cite{Esther2015} demonstrated the effectiveness of a feature selection technique based on support vector machines (SVMs) that uses the weight vector to categorize dementia in high-dimensional neuroimaging data.
Jie et al. \cite{jie2018discriminating} used whole-brain FC and FOBA with SVM to distinguish bipolar disorder from serious depression. Using a majority-voting method, Holker and Susan \cite{holker2021quantitative} ranked EEG characteristics that were important for detecting alcohol use disorders. Ali et al. \cite{ali2023correlation} introduced a correlation-filter-based method for selecting informative channels in EEG-fNIRS systems within a hybrid architecture. Using infinite feature selection, authors \cite{obertino2016infinite} were able to uncover changes in connection following a stroke. In the study \cite{pei2020eeg}, the authors utilized a hybrid method using feature fusion and selection to identify workloads based on EEG data.
Functional connection in clinical neuroscience is shown by the wide range of illnesses and methods in the literature. Most of these studies favor classification accuracy over feature stability and reproducibility, which is crucial for reliable biomarkers. Our work assesses the discriminative strength and consistency of selected features across cross-validation folds using Kuncheva and Jaccard indices. This dual goal improves explainable AI brain illness diagnosis interpretability and reliability.


\section{Materials and Method}
\label{sec:method}
In this paper, we have used a vectorized connectome as our input feature vector for our model. As the number of features is much larger than the data samples, effective feature selection must be used. Here we have evaluated feature selection techniques like embedded feature selection, filter-based feature selection, and wrapper-based feature selection. We have used logistic regression (LR) as our backbone classifier for classifying Schizophrenia from healthy patients.


\subsection{Datasets}
Here we have used two types of datasets, for our experiment, the descriptions of which are given below:
\subsubsection{Dataset 1}
The dataset used in our experiments is downloaded from \cite{Vohryzek2020}.
The cohort consists of a schizophrenic group of 27 subjects and a control group of 27 healthy subjects. All of the data was collected at the Service of General Psychiatry at the Lausanne University Hospital. The schizophrenic patients were diagnosed with schizophrenia and schizoaffective disorders after meeting the DSM-IV criteria. Control subjects had no history of neurological disease. All 54 subjects had given written consent following the institutional guidelines approved by the Ethics Committee of Clinical Research of the Faculty of Biology and Medicine, University of Lausanne, Switzerland.

A 3 Tesla Siemens Trio scanner equipped with a 32-channel head coil was used to acquire imaging data from all participants. Functional imaging was performed using a gradient echo planar imaging (EPI) sequence optimized for BOLD contrast, with an in-plane resolution and slice thickness of 3.3 mm, along with a 0.3-mm gap. The echo time (TE) was 30 ms, and the repetition time (TR) was 1,920 ms, resulting in 280 images per subject.
Functional connectomes were derived from fMRI BOLD time-series by computing the absolute Pearson correlation between the time-courses of individual brain regions. Here we have considered 83 regions of interest(ROI) for our study.

\begin{figure*}
    \centering
    \subfigure[Healthy.]{\includegraphics[width=0.4\textwidth]{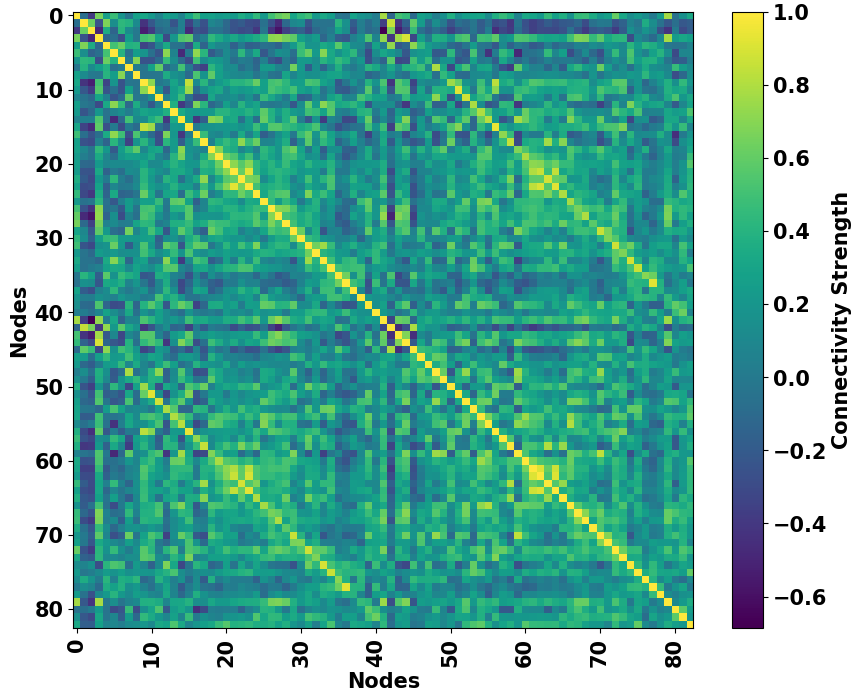}} 
    \subfigure[Diseased.]{\includegraphics[width=0.4\textwidth]{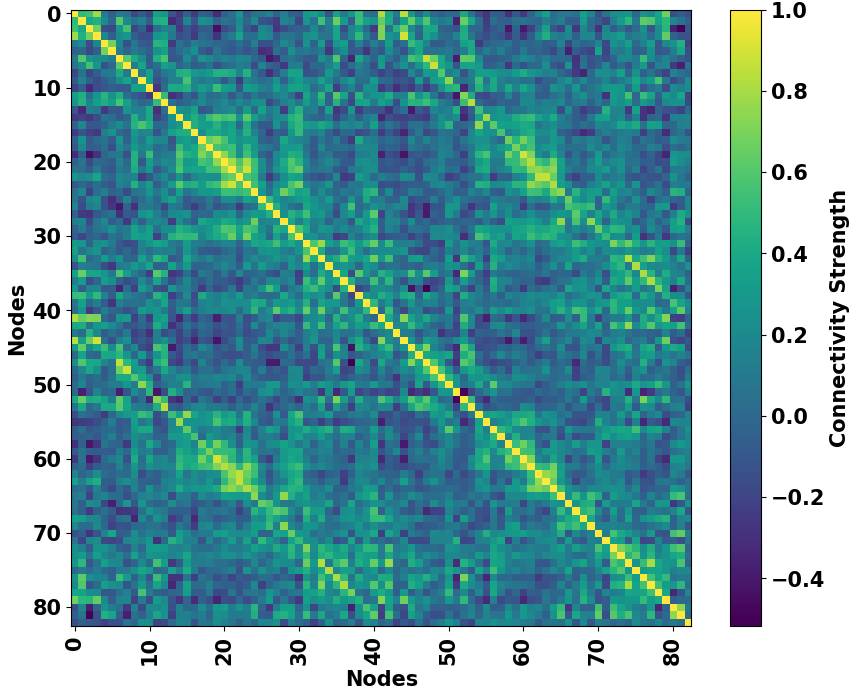}}
    \caption{Example of functional connectomes of healthy and diseased subjects from Dataset 1 (provided by University of Lausanne, Switzerland). A symmetric matrix shows the pairwise Pearson correlation between 83 brain areas in each subject's connectome. The matrix values are thresholded absolute correlations; warmer colors indicate greater functional relationships.} 
    \label{fig:dataset1}
\end{figure*}

\begin{figure*}
    \centering
    \subfigure[Healthy.]{\includegraphics[width=0.4\textwidth]{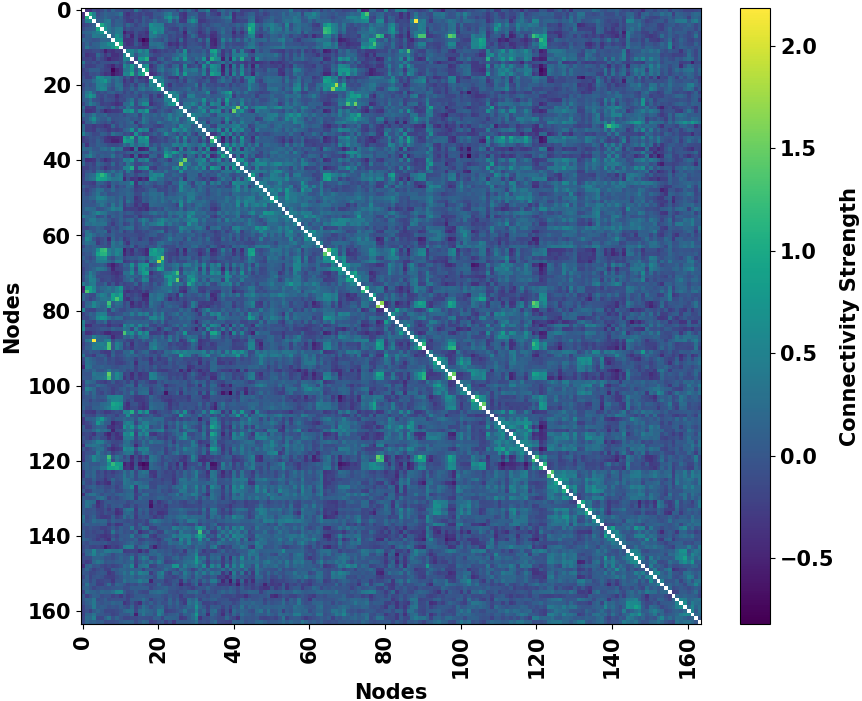}} 
    \subfigure[Diseased.]{\includegraphics[width=0.4\textwidth]{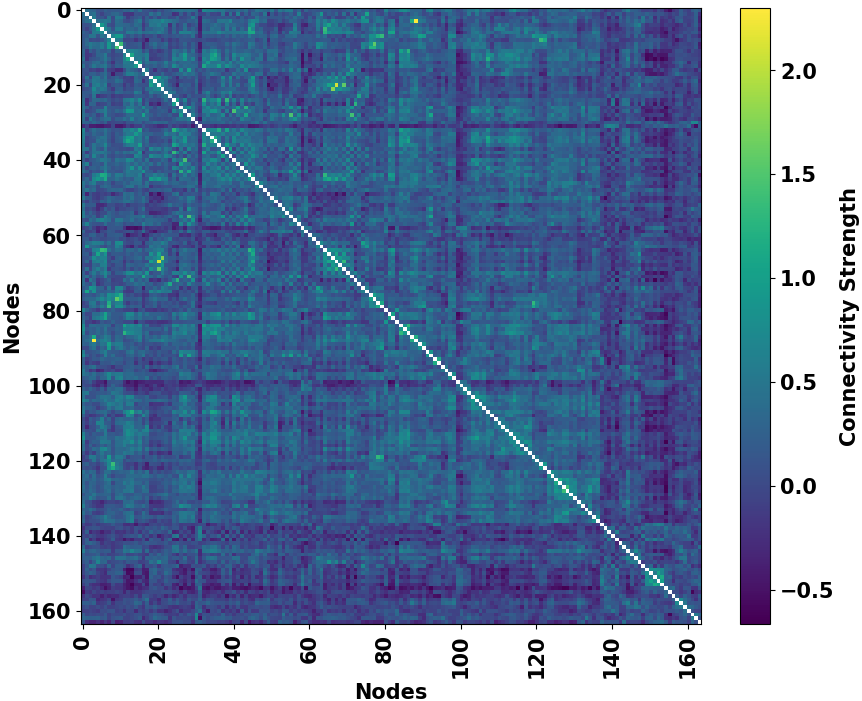}}
    \caption{Example of functional connectomes of healthy and diseased subjects from Dataset 2 (provided by UCLA, USA). The matrices denote the pairwise association among 164 brain regions. These high-resolution connectomes reveal more intricate inter-regional interactions than Dataset 1. The visual disparities between the two groups demonstrate fundamental abnormalities in connection patterns pertinent to classification.}
    \label{fig:dataset1}
\end{figure*}

\begin{table}
\centering
\renewcommand{\arraystretch}{1.5}
\begin{tabular}{|l|c|} 
\hline
\textbf{Symbols} & \textbf{Significance } \\
\hline
$i$  & Subject index\\  
$n$ & Total number of subjects.  \\

$x_i$ & Input feature of $i$-th subject. \\

$d$& Number of ROI's\\
$p = \frac{d(d-1)}{2}$ & Length of each feature vector.\\
$x_{ij}$, $j = 1, 2, \ldots, p$ & $j$-th feature component of $i$-th subject. \\
$y_i \in \{0, 1\}$ & Class of $i$-th subject. \\

- - - - - - - - - - & - - - - - - -  - - - - -  - - - - - - - - - - \\
$w \in \mathbb{R}^{q \times 1}$ & Weight vector in LR computation. \\
$b$ & Intercept parameter in LR computation.\\
- - - - - - - - - - & - - - - - - -  - - - - -  - - - - - - - - - - \\
$\alpha$ & Intercept parameter in LASSO computation.\\
$\beta$ & Weight assignment vector of length $p$ in LASSO.\\
$\lambda$ & Regularization parameter in LASSO.\\
$v \in \mathbb{R}^{p \times 1}$ & Weight assignment vector of length $p$ in Relief. \\
$F_j$& F-Score for $j$-th feature using ANOVA.\\

\hline
\end{tabular}
\caption{Summary of the variables and definitions used in this paper.}
\label{tab:symbol}
\end{table}


\subsubsection{Dataset 2}
This dataset consists of fMRI data acquired at the University of California, Los Angeles (UCLA) Consortium for Neuropsychiatric Phenomics LA5c study, which contains 50 schizophrenic subjects and 122 control subjects \cite{poldrack2016phenome}. Each fMRI data set, having $60 \times 60 \times 30 = 108, 000$ voxels is preprocessed through: functional realignment, slice timing correction, outlier detection, segmentation, normalization, functional smoothing, and temporal bandpass filtering. The brain is parcellated into 164 regions-of-interest (ROI) in accordance with AAL3 Atlas \cite{rolls2020automated} and Harvard-Oxford Cortical Atlas \cite{kennedy2016harvard}. The connectivity between each pair of ROIS is measured in terms of  Pearson correlation coefficient (PCC) \cite{fox2009global,murphy2009impact}.
The finally-processed functional connectomes are publicly available from Sunil et al. \cite{Sunil2024}. 

\subsection{Logistic Regression}
Here we have used logistic regression \cite{DREISEITL2002352} as our backbone classifier. Because it works well when the dataset is small to moderately sized and doesn’t require large computational resources \cite{KURT2008366}. Given a set of data examples $\mathbf{x}_i \in \mathbb{R}^p, i = 1, \ldots, n$, and a vector $\mathbf{y} \in \{0, 1\}^n$ representing the group membership of data points, logistic regression aims to model the probability that each data point belongs to a particular class. The logistic regression model finds the best-fitting hyperplane that separates the two classes by maximizing the likelihood of the observed data. The mathematical formulation can be written as a maximum likelihood estimation problem:

\begin{equation} 
\begin{split}
\max_{\mathbf{w}, b} \sum_{i=1}^n & \big( y_i \log(\sigma(\mathbf{w}^T \mathbf{x}_i + b)) \\
& + (1 - y_i) \log(1 - \sigma(\mathbf{w}^T \mathbf{x}_i + b)) \big),  
\end{split}
\end{equation}

\noindent
where $\sigma(z) = \frac{1}{1 + e^{-z}}$ is the logistic function, $\mathbf{w} \in \mathbb{R}^p$ is a vector of coefficients, and $b$ is the intercept term. A classifier that generalizes well is then found by controlling the complexity of the model, often by regularizing $\mathbf{w}$ to prevent overfitting, typically using $L_2$ regularization. For the sake of clarity, we have summarized the symbols in Table \ref{tab:symbol}.


\subsection{Feature selection algorithms}

\subsubsection{LASSO}
In this work, we utilize the least absolute shrinkage and selection operator (LASSO) as our method for both regularization and feature selection. LASSO \cite{10.1111/j.2517-6161.1996.tb02080.x} is particularly effective in scenarios with high-dimensional data, where the number of features might exceed the number of observations.

LASSO modifies the ordinary least squares objective by adding a regularization term that is the \( L_1 \) norm of the coefficients. This addition encourages sparsity in the coefficients of the model, effectively performing feature selection by driving some coefficients to exactly zero. The mathematical formulation for LASSO is given by an unconstrained optimization problem:
\begin{equation}
\min_{\boldsymbol{\beta}} \left\{ \frac{1}{2n} \sum_{i=1}^n (y_i - \beta_0 - \sum_{j=1}^p \beta_j \mathbf{x}_{ij})^2 + \lambda \sum_{j=1}^p |\beta_j| \right\}
\label{eq:lasso}
\end{equation}

The choice of $\lambda$ in \eqref{eq:lasso} is critical in LASSO as it determines the level of regularization; higher values lead to greater regularization. Typically, the regularization parameter is selected through a cross-validation approach to balance between model complexity and data fidelity.


\subsubsection{Relief}
Given a randomly selected subject $x_l$, we first find the \emph{hit set} $\mathcal{H}$ and \emph{miss set} $\mathcal{M}$ of it. For example, $\mathcal{H}$ is the collection of the rest of the healthy subjects if the randomly selected subject $x_l$ is healthy. Similarly, $\mathcal{M}$ is the collection of all diseased subjects if the subject $x_l$ is diseased. 

\begin{equation}
    x_l^{hit} = \underset{x_i \in \mathcal{H}}{\text{argmin}} || x_l - x_i||^2. 
\end{equation}

\begin{equation}
    x_l^{miss} = \underset{x_i \in \mathcal{M}}{\text{argmin}} || x_l - x_i||^2. 
\end{equation}

We initialize the  weights $v_j$ for  $j=1, 2, \ldots, p$ with zeros. Then the aggregate weight for the  $j$-th feature is given by
\begin{equation}
    v_j = v_j - (x_{lj} - x_{lj}^{hit})^2 + (x_{lj} - x_{lj}^{miss})^2,
\end{equation}
where $x_{lj}$ is the $j$-th feature component of $l$-th subject, $x_{lj}^{hit}$ is the $j$-th feature component of nearest neighbor from it's hit set $\mathcal{H}$, and  $x_{lj}^{miss}$ is the $j$-th feature component of Euclidean distance nearest neighbor from it's miss set $\mathcal{M}$.

Notice that the weight of any given feature decreases if it differs from that feature in nearby instances of the same class more than in nearby instances of the other class, and increases in the reverse case. After a certain number of iterations, divide each element of the weight vector by the number of iterations. This becomes the relevance vector. There is no hard stopping condition based on convergence or accuracy; instead, the number of iterations is a hyperparameter that controls how well the feature relevance estimates generalize. 

\subsubsection{Analysis of Variance (ANOVA)}
Analysis of variance (ANOVA) compares the variance between groups (i.e., different classes) and the variance within groups (i.e., samples in the same class). The goal is to determine whether a feature’s values vary significantly across different classes. The key steps are as follows:

\begin{enumerate}
    \item Calculate the mean of each class: For a given $j$-th feature, the mean value for each class is computed as:
    
\begin{equation}
\begin{split}
     & x_{0j}^{H} = \frac{1}{n_H} \sum_{i=1}^{n_H} x_{ij}, \quad
    x_{1j}^{D} = \frac{1}{n_D} \sum_{i=1}^{n_D} x_{ij}, \\
    & \overline{x}_{jj} = \frac{1}{n} \sum_{i=i}^{n} x_{ij},   \quad n = n_H + n_D,
\end{split}
    \label{eq:class_mean}
\end{equation}

where
\begin{itemize}
    \item $n_H$ is the number of healthy subjects (class $0$).
    \item $n_D$ is the number of diseased subjects (class $1$).

    \item $n$ is the total number of subjects. 

    \item $x_{0j}^{H}$ is the mean of $j$-th feature-component for class $0$ which is of healthy subjects.
    \item $x_{1j}^{D}$ is the mean of $j$-th feature-component  for class $1$ which is of diseased subjects.

    \item $\overline{x}_{jj}$ is the mean $j$-th feature-component  for all subjects.
\end{itemize}

\item Compute between-class variance (SSB): 

The variance between classes, for the $jth$ feature, also called the Sum of Squares Between (SSB), is given by:

\begin{equation*}
SSB_j =  n_H (x^{H}_{0j} - \overline{x}_{jj})^2 + n_D (x^{D}_{1j} - \overline{x}_{jj})^2
\label{eq:ssb}
\end{equation*}
For binary classification, we have mean square between groups (MSB) for the $j$-th component of the feature vector is: 
\begin{equation*}
MSB_j = SSB_j.
\label{eq:msb}
\end{equation*}
\item Compute within-class variance: For the $j$-th feature, the variance within each class, referred to as, sum of squares within (SSW), is:

\begin{equation*}
SSW_j =  \sum_{i=1}^{n_H} (x_{ij} - x^{H}_{0j})^2 +  \sum_{i=1}^{n_D} (x_{ij} - x^{D}_{1j})^2 
\label{eq:ssw}
\end{equation*}
For binary classification, we have the \texttt{mean square within groups} (MSW) for $j$-th component of feature vector as:
\begin{equation*}
MSW_{j} = \frac{SSW_{j}}{n-1}
\label{eq:msw}
\end{equation*}

\item Compute F-statistic: Finally, the F-statistic for the $j$-th component of the feature vector is calculated as follows:
\begin{equation}
F_j = \frac{MSB_j}{MSW_j}.
\label{eq:f_final}
\end{equation}
\end{enumerate}
A higher $F_j$ value indicates that the feature is more useful for classification.


\subsection{Stability Analysis}

The stability of selected features is a critical aspect in  
feature selection tasks. It measures how consistently the same
features are selected across different iterations, datasets, or
models. A stable feature selection method indicates robustness
and reliability. Here we have used the Kuncheva stability index
and the Jaccard Index to assess the stability of the features that
are selected by the algorithms.


\subsubsection{Kuncheva Index}
The Kuncheva index (KI) \cite{Kuncheva2007} is a statistical measure used to assess the stability and consistency of feature selection across different subsets of data in machine learning and data analysis.
In our experiments, we examine the stability of total $p$ components in an FC feature of a particular subject. We exclude self-loops, which means that the diagonal non-zero values of the connectivity matrices are discarded.

Suppose that two subsets of feature components are $j_{\theta}$ and $j_{\phi}$. Each subset has $k$ elements. The total number of components in the feature vector of each subject is $p$. The larger the intersection between the subsets, the higher the value of the consistency in selection of the feature-component. Let $r$ be the cardinality of the intersection of the two subsets.
The maximum $r$ equals $k$ is achieved when both $j_{\theta}$ and $j_{\phi}$ are identical subsets. The Kuncheva index (KI) \cite{Kuncheva2007} is given by:

\begin{equation}
K(j_{\theta}, j_{\phi}) = \frac{r - (k^2/p)}{k - (k^2/p)},
\label{eq:Kuncheva}
\end{equation}
where \( r = |j_{\theta} \cap j_{\phi}| \) and \( k = |j_{\theta} | = |j_{\theta} | \), the size of each subset of feature components. 

In our experiments, we draw $\lambda = 100$ subsamplings to obtain a respective feature signature, i.e. sequence of indices of selected features.
The overall stability index \cite{Kuncheva2007} for a sequence of signatures is defined as the average of all pairwise stability indices on \( k \) subsamplings:
\begin{equation}
K_{\text{avg}} = \frac{2}{\lambda (\lambda  - 1)} \sum_{\theta = 1}^{\lambda  - 1} \sum_{\phi = \theta+1}^{\lambda } K ( j_{\theta}, j_{\phi}).
\end{equation}
Note that a higher value of $K$ in \eqref{eq:Kuncheva} indicates a higher stability.

\subsubsection{Jaccard Index}
Jaccard’s index (JI) \cite{mohana2016survey} measures the average similarity from all pairwise selected feature subsets:
\begin{equation}
J_0 (j_{\theta}, j_{\phi}) = \frac{|j_{\theta} \cap  j_{\phi}|}{|j_{\theta} \cup j_{\phi}|}
\label{eq:JI}
\end{equation}

\begin{equation}
J_{\text{avg}} = \frac{2}{\lambda (\lambda  - 1)} \sum_{\theta = 1}^{\lambda  - 1} \sum_{\phi = \theta+1}^{\lambda } J ( j_{\theta}, j_{\phi}).
\end{equation}

The stability index $J_0$ in \eqref{eq:JI} lies in the range $[0, 1]$. A high value close to $1$ implies the algorithm is stable \cite{mohammadi2016robust}. 

\section{Experimental Results}
\label{sec:results}

\begin{figure*}
    \centering
    \subfigure[Accuracy.]{\includegraphics[width=0.4\textwidth]{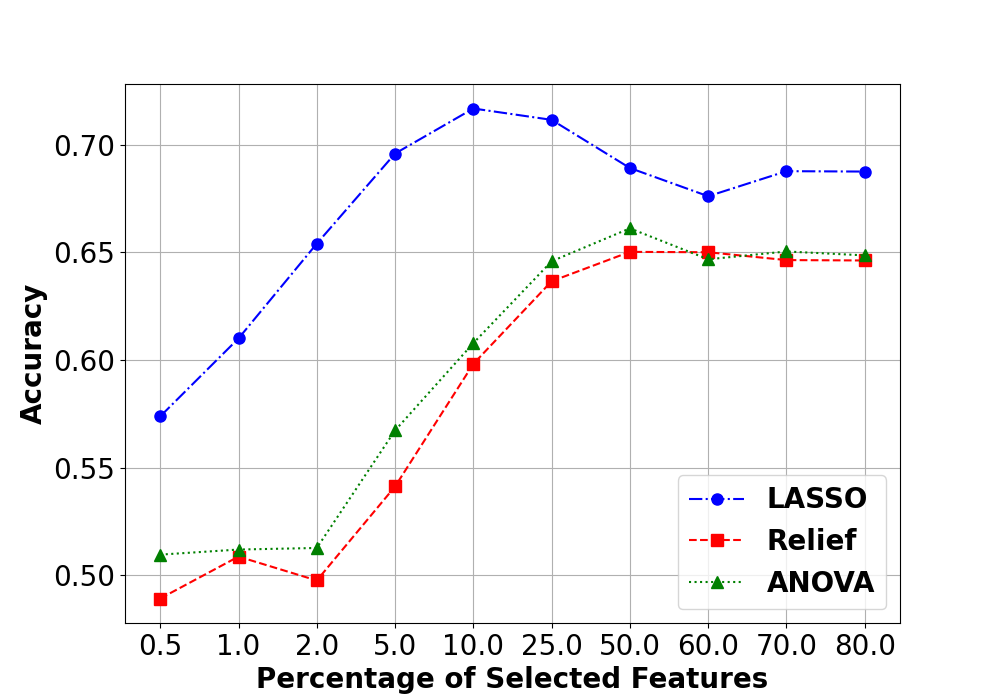}} 
    \subfigure[F1 score.]{\includegraphics[width=0.4\textwidth]{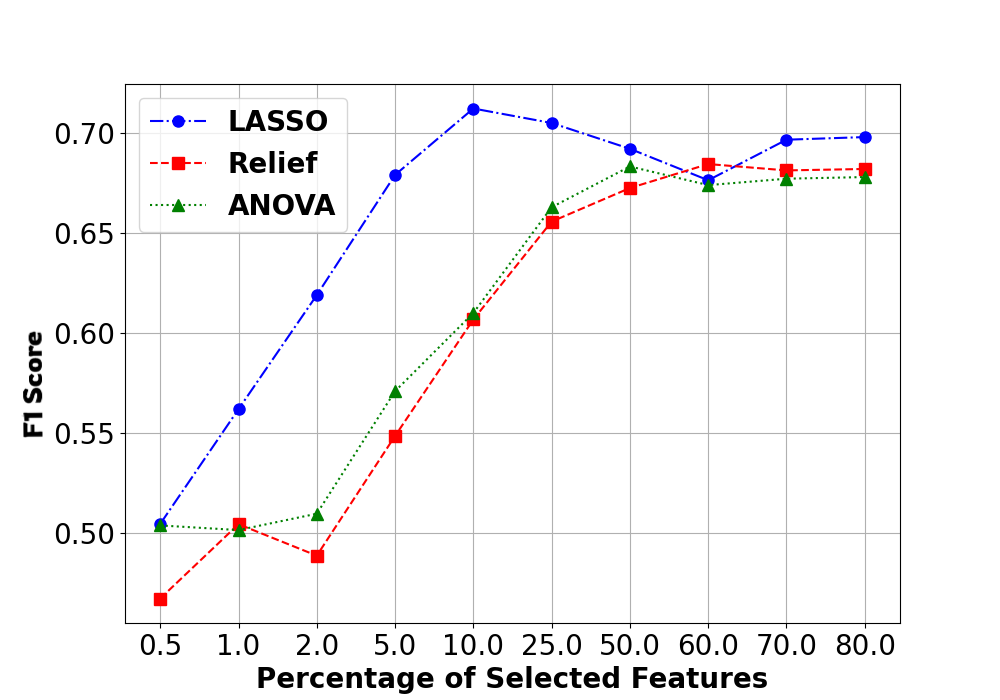}} \\
    \subfigure[Kuncheva index.]{\includegraphics[width=0.4\textwidth]{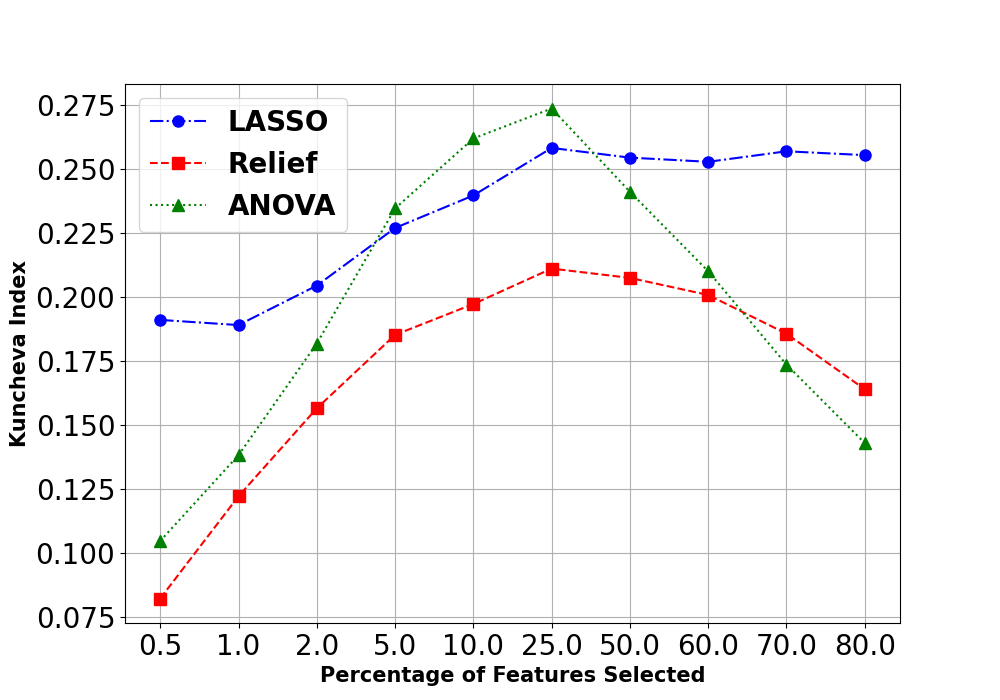}} 
    \subfigure[Jaccard index.]{\includegraphics[width=0.4\textwidth]{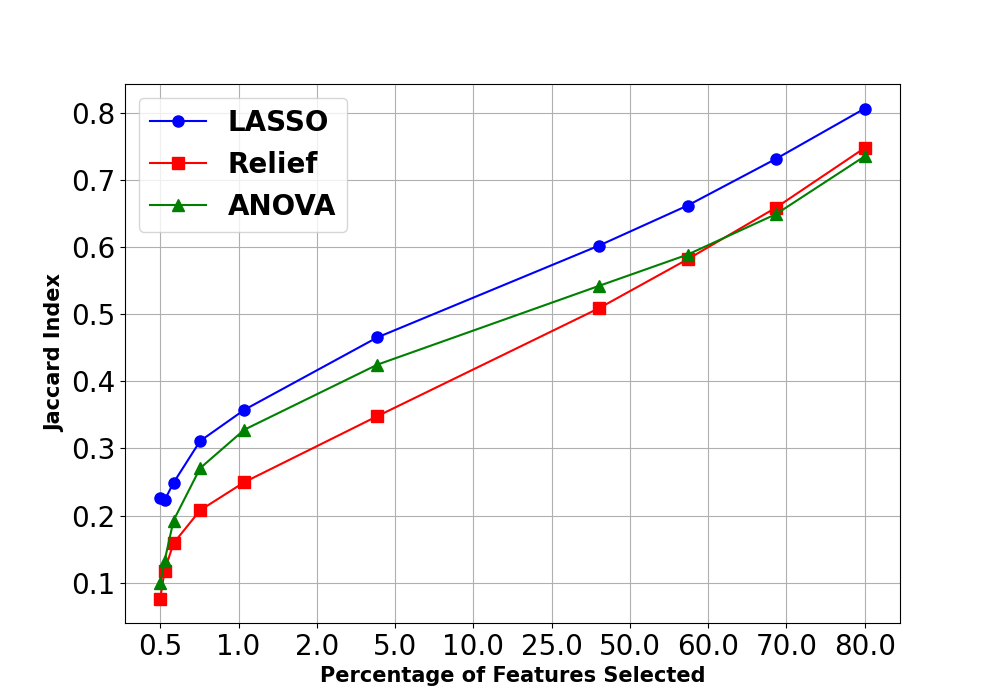}}
    \caption{Classification performance and feature selection stability on Dataset 1 (University of Lausanne). This dataset includes 54 functional connectomes vectorized into 3,403 characteristics from 27 healthy controls and 27 schizophrenic patients. LASSO, Relief, and ANOVA are tested at different percentages of selected characteristics in subfigures. LASSO consistently surpasses others in classification accuracy and F1 score. Kuncheva index and Jaccard index assess feature stability across 100 cross-validation subsamplings. LASSO has the highest stability across both criteria, indicating more reproducible biomarker selection.}
    \label{fig:dataset1}
\end{figure*}

\subsection{Experimental Settings}

\subsubsection{Nested 5-Fold Cross-Validation Framework}

The process utilizes two levels of cross-validation: the external loop for unbiased performance estimation and the internal loop for model tuning and parameter optimization \cite{GUTIERREZGOMEZ2020102316}. 
We use the cross-validation experimental scheme as in \cite{GUTIERREZGOMEZ2020102316}.

\textbf{External Loop:} The dataset is divided into 5 folds. This external CV loop is shuffled 20 times, generating 100 subsamplings of the original dataset. In each shuffle, 80\% of the data (four folds) serves as the training set for the internal CV, while the remaining 20\% (one fold) acts as the testing set to provide an unbiased evaluation of the model.

\textbf{Internal  Loop}: Within the training set of the external CV, the internal CV loop is applied to further split the data into training and validation subsets. Four folds are used as the training set, and one fold (held-out) is used as the validation set to tune model parameters and select features.

\begin{figure*}
    \centering
    \subfigure[Accuracy.]{\includegraphics[width=0.41\textwidth]{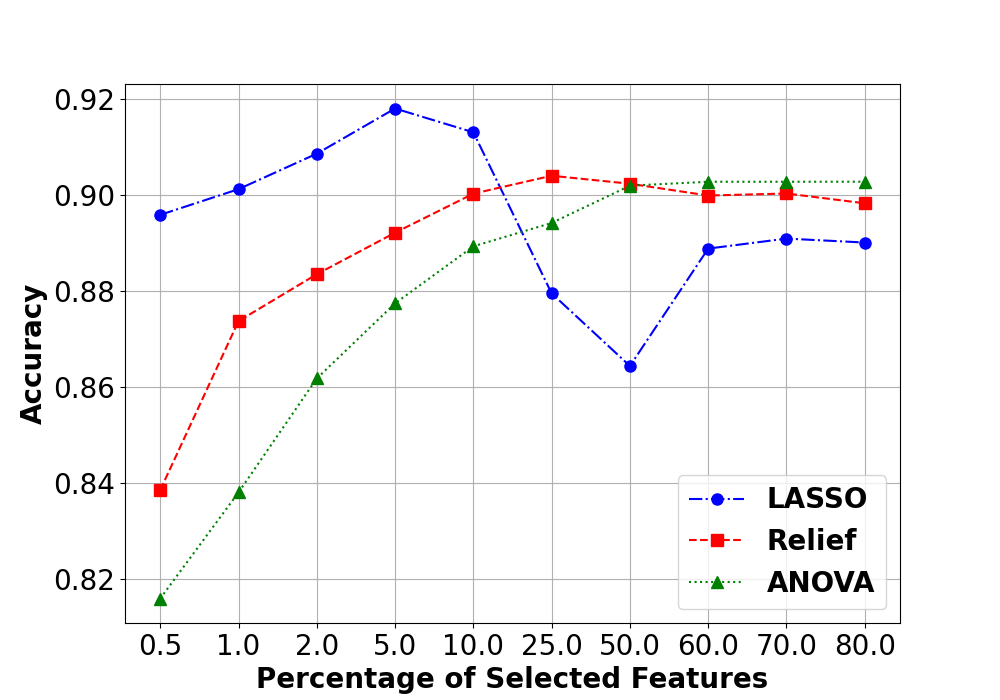}} 
    \subfigure[F1 score.]{\includegraphics[width=0.41\textwidth]{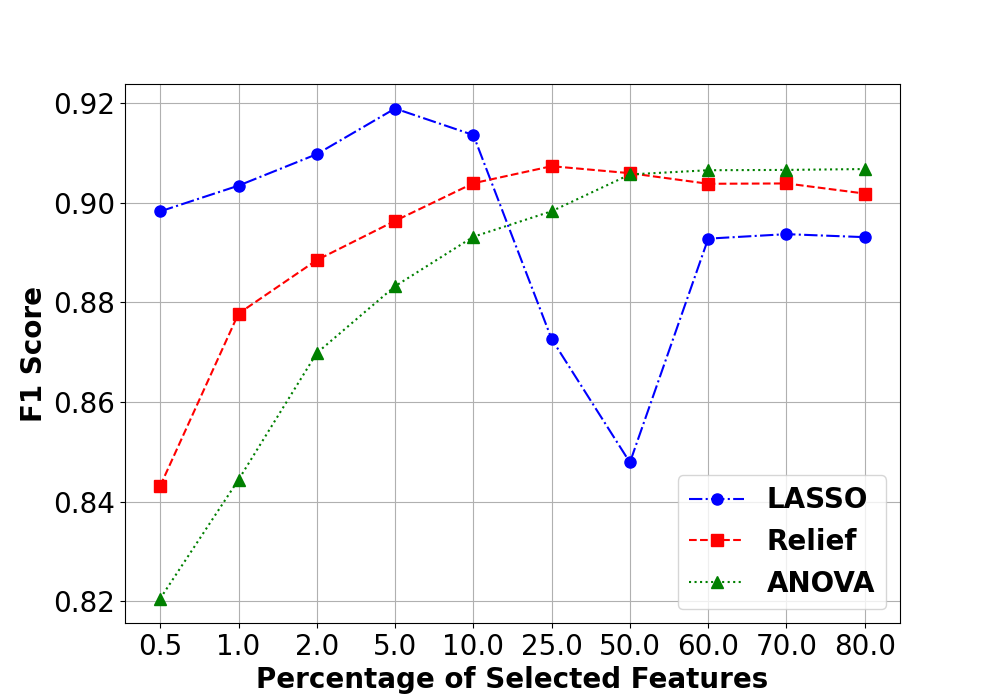}}\\
    \subfigure[Kuncheva index.]{\includegraphics[width=0.41\textwidth]{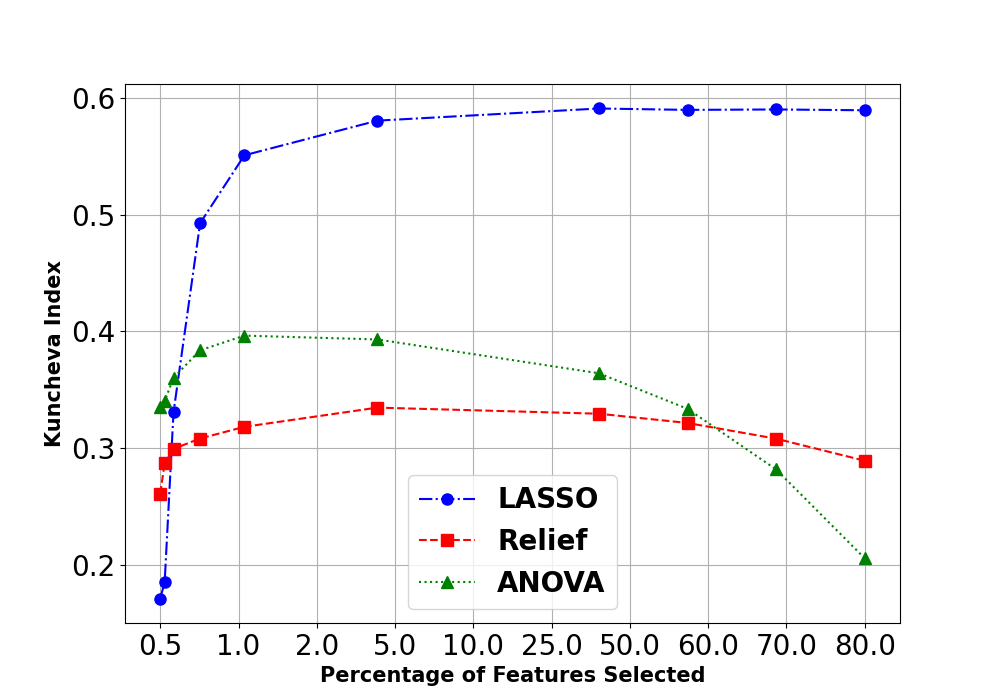}} 
    \subfigure[Jaccard index.]{\includegraphics[width=0.41\textwidth]{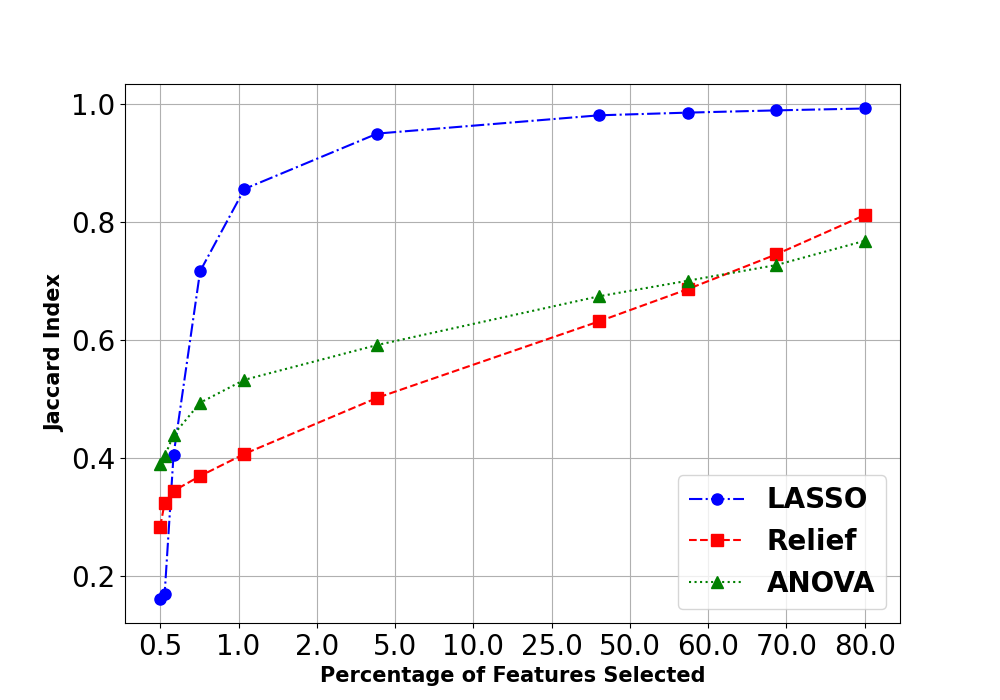}}
    \caption{Classification performance and feature selection stability on Dataset 2 (UCLA). The functional connectomes of 172 people (122 healthy controls and 50 schizophrenia patients) were vectorized into 13,366 features from 164 brain areas.  The feature selection methods LASSO, Relief, and ANOVA are tested at different feature percentages.  LASSO consistently outperforms in classification accuracy and F1 score.  Kuncheva index and Jaccard index evaluate feature stability across 100 cross-validation subsamplings.  LASSO has the strongest stability across both criteria, proving its reproducibility for large-scale connectome biomarker detection.}
    \label{fig:dataset2}
\end{figure*}

\subsubsection{Data Augmentation}
We have used the synthetic minority oversampling technique (SMOTE) \cite{chawla2002smote} for augmenting the data in Dataset 2. The main principle behind SMOTE is, it selects samples from the minority class, determines the k-nearest neighbors, and finally creates a new sample by interpolating between a selected sample and one of its neighbors:
\begin{equation}
    x_{\text{new}} = x_{\text{original}} + \lambda  (x_{\text{neighbor}} - x_{\text{original}})
\end{equation}
where $\lambda$ is a random number between 0 and 1. The above data augmentation step is for class balancing in Dataset 2.

\begin{table}
\centering
\renewcommand{\arraystretch}{1.5}
\begin{tabular}{|l|c|c|c|c|} 
\hline
\textbf{Methods} & \textbf{ Accuracy } & \textbf{Precision} & \textbf{Recall} & \textbf{F1-Score} \\
\hline
LASSO  & 70 &  72.40  &  67.27  &  66.42  \\  
Relief & 54.12 &  54.01  &  58.27  &  55  \\ 
ANOVA  & 56.32 &  57.43  &  60.36  &  56.25  \\
\hline
\end{tabular}
\caption{Dataset 1. Classification using top $5\%$ components in FC.}
\label{tab:dataset01_05}
\end{table}

\begin{table}
\centering
\renewcommand{\arraystretch}{1.5}
\begin{tabular}{|l|c|c|c|c|} 
\hline
\textbf{Methods} & \textbf{Accuracy } & \textbf{Precision} & \textbf{Recall} & \textbf{F1-Score} \\
\hline
LASSO  & 71.67 &  73.49  &  71.67  &  71.26  \\  
Relief & 60 &  59.46  &  64.87  &  60.56  \\ 
ANOVA  & 60.55 &  61.74  &  64.47  &  61.01  \\   
\hline
\end{tabular}
\caption{Dataset 1. Classification using top $10\%$ components in FC.}
\label{tab:dataset01_10}
\end{table}

\subsection{Results on Classification of Brain Disorders}

In this section, we present a detailed analysis of the performance metrics derived from various feature selection techniques applied within our study. In particular, we evaluate the accuracy of the models that incorporate these feature selection methods: LASSO, Relief, and ANOVA.
To quantitatively measure the accuracy, we plotted the classification accuracies for each feature selection method. This allowed us to compare how each method impacts the predictive performance of the models. The graphical representations were designed to provide a clear and comparative perspective of the efficacy of each feature selection approach under varying conditions.

\subsubsection{Dataset 1. University of Lausanne:} This dataset has functional connectomes of 27 healthy subjects and 27 schizophrenia (diseased) patients.  Each connectome has $d = 83$ brain regions; therefore, $\frac{ 83 \times (83-1)}{2} = 3,403$ components in the vectorised connectome of each subject. We first study the classification performance by three feature selection for different percentage of selected components shown in Fig. \ref{fig:dataset1}. Notice in Fig. \ref{fig:dataset1}(a) and \ref{fig:dataset1}(b) that LASSO selection offers better accuracy and F1 score respectively than both Relief and ANOVA.  In Table \ref{tab:dataset01_05}, we report four metrics - accuracy, precision, recall, and F1 score when top $5\%$  components are selected (around 170 from a total of 3403) to be input to the logistic regression classifier. Similarly, we report the results for top $10\%$  components in Table \ref{tab:dataset01_10}. 

\begin{figure*}
    \centering
        \subfigure[LASSO.]{\includegraphics[width=0.32\textwidth]{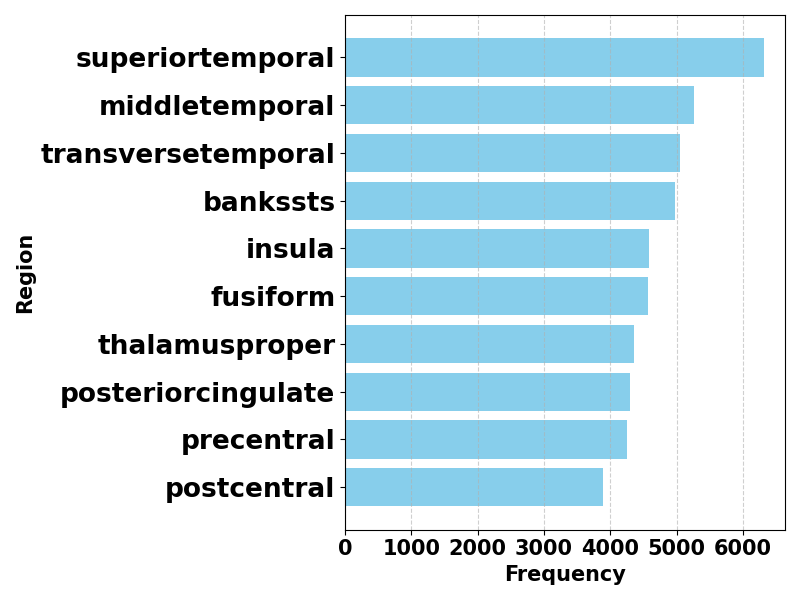}}
        \subfigure[ANOVA.]{\includegraphics[width=0.32\textwidth]{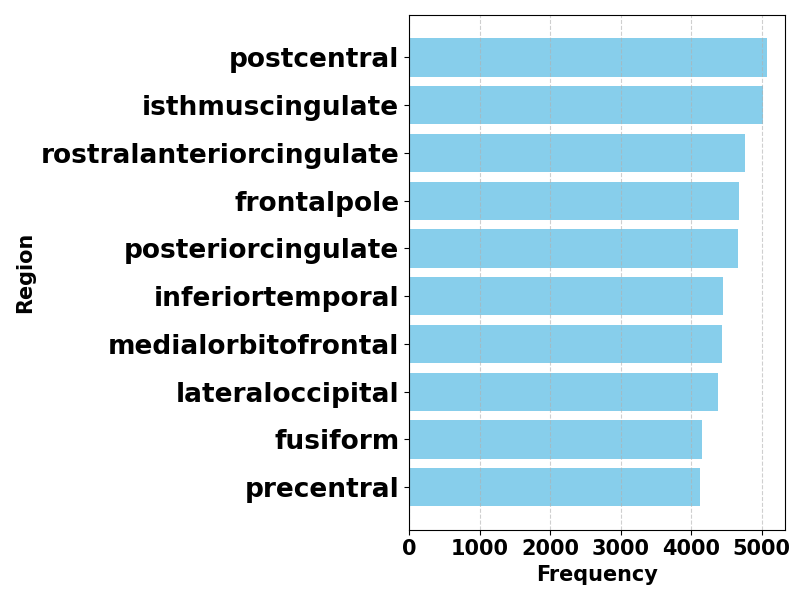}}  
        \subfigure[Relief.]{\includegraphics[width=0.32\textwidth]{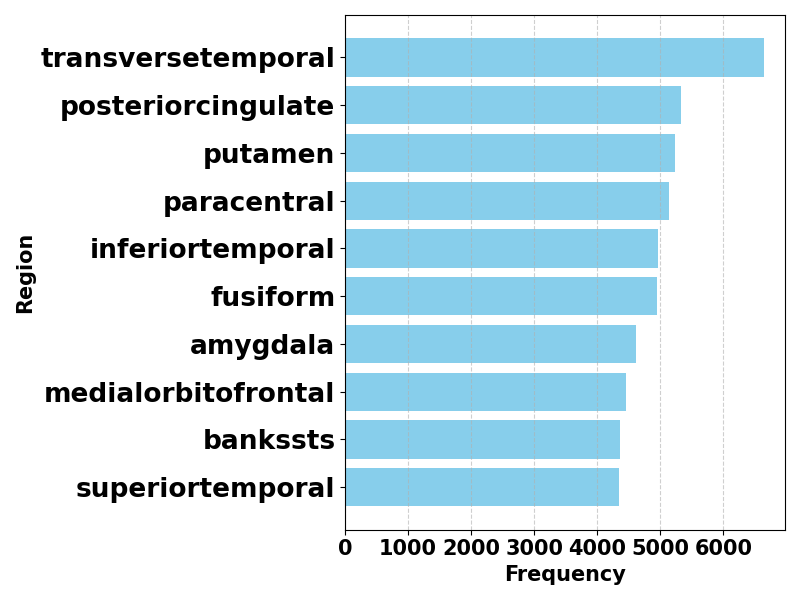}} \\
        \texttt{\textbf{Top 10 ROIs selected in left-hemisphere.}} \\ \vspace{5mm}
        \subfigure[LASSO.]{\includegraphics[width=0.32\textwidth]{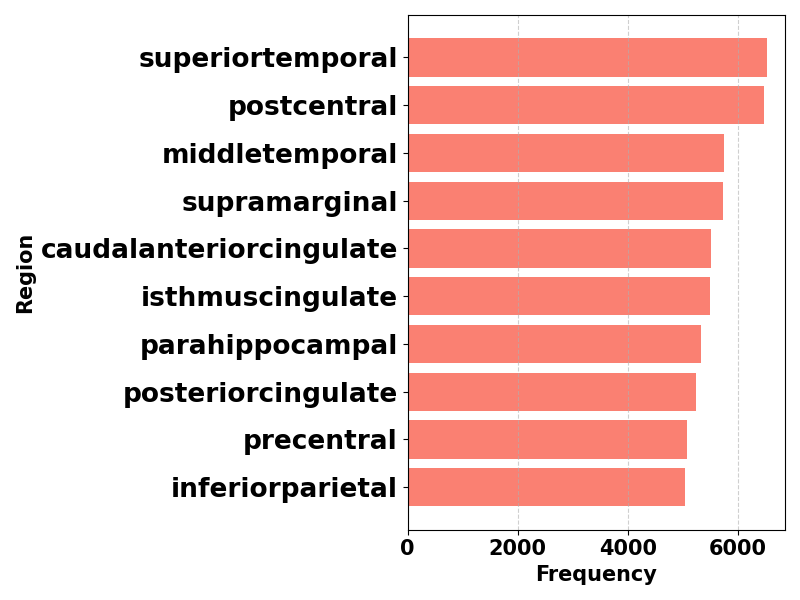}}
        \subfigure[ANOVA.]{\includegraphics[width=0.32\textwidth]{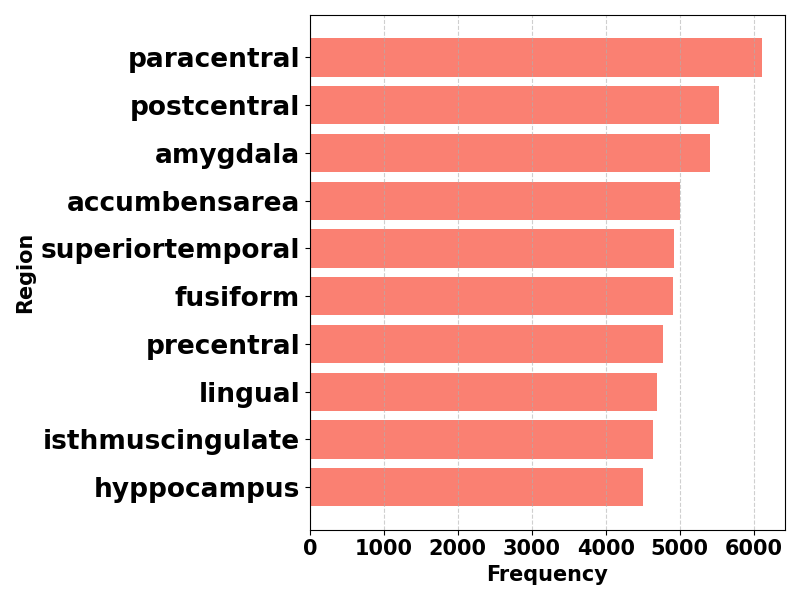}}
        \subfigure[Relief.]{\includegraphics[width=0.32\textwidth]{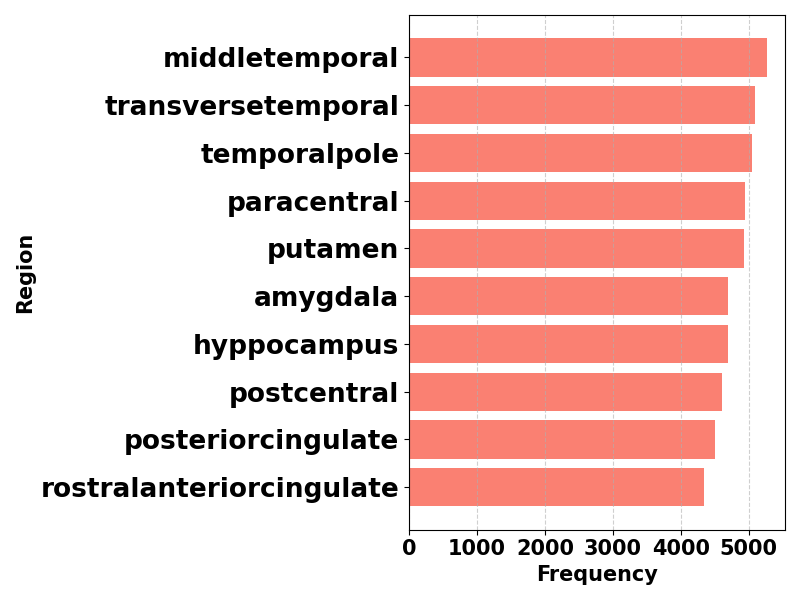}}   \\
        \texttt{\textbf{Top 10 ROIs selected in right-hemisphere.}}
    \caption{Affected regions for dataset 1. In our experiment, we have 5-fold cross validation in each trial consisting of 7 different values of percentage of selected feature-component: $[0.5, 1, 2, 5, 10, 25, 50]$. We repeated the experiment over 10 trials. Therefore, for a particular brain region, the maximum possible occurrence will be: $10 \times 7 \times 5 (d-1) = 350 \times 82 = 28700$.}
    \label{fig:affected_ds1}
\end{figure*}

\subsubsection{Dataset 2. UCLA:} This dataset has functional connectomes of 122 healthy subjects and 50 schizophrenia (diseased) patients.  Each connectome has $d = 164$ brain regions; therefore, $\frac{ 164 \times (164-1)}{2} = 13,366$ components in the vectorised connectome of each subject. We first study the classification performance by three feature selection for different percentage of selected components shown in Fig. \ref{fig:dataset2}. Notice in Fig. \ref{fig:dataset2}(a) and \ref{fig:dataset2}(b) that LASSO selection offers better accuracy and F1 score respectively than both Relief and ANOVA when top $10\%$ components are selected.  In Table \ref{tab:dataset02_05}, we report four metrics - accuracy, precision, recall, and F1 score when top $5\%$  components are selected (around 668 from a total of 13,366) to be input to the logistic regression classifier. Similarly, we report the results for the top $10\%$  components in Table \ref{tab:dataset02_10}. 

In terms of classification accuracy and F1 score, LASSO consistently outperformed Relief and ANOVA across both datasets. LASSO used the top 10\% of functional connectivity features to reach a peak accuracy of 71.67\% and an F1 score of 71.26\% for Dataset 1 (Tables \ref{tab:dataset01_05} and \ref{tab:dataset01_10}). With F1 values of 60\% or below, relief and ANOVA fell behind. Using the top 5\% of features, LASSO achieved 91.85\% classification accuracy and 91.98\% F1 score (Table \ref{tab:dataset02_05}); minimal variations were seen when increasing to 10\% (Table \ref{tab:dataset02_10}). These outcomes highlight the benefit of utilizing integrated techniques that simultaneously optimize feature selection and classification, such as LASSO.

\begin{table}
\centering
\renewcommand{\arraystretch}{1.5}
\begin{tabular}{|l|c|c|c|c|} 
\hline
\textbf{Methods} & \textbf{ Accuracy } & \textbf{Precision} & \textbf{Recall} & \textbf{F1-Score} \\
\hline
LASSO  & 91.85 &  92.22  &  91.79  &  91.98  \\  
Relief & 89.25 &  87.73  &  92.02  &  88.89  \\ 
ANOVA  & 87.95 &  85.58  &  91.61  &  88.98  \\
\hline
\end{tabular}
\caption{Dataset 2. Classification using top $5\%$ components in FC.}
\label{tab:dataset02_05}
\end{table}
\begin{table}
\centering
\renewcommand{\arraystretch}{1.5}
\begin{tabular}{|l|c|c|c|c|} 
\hline
\textbf{Methods} & \textbf{Accuracy } & \textbf{Precision} & \textbf{Recall} & \textbf{F1-Score} \\
\hline
LASSO  & 91.25 &  91.75  &  91.22  &  91.56  \\  
Relief & 90 &  88.48  &  92.67  &  90.56  \\ 
ANOVA  & 89.12 &  87.43  &  91.61  &  89.14  \\  
 
\hline
\end{tabular}
\caption{Dataset 2. Classification using top $10\%$ components in FC.}
\label{tab:dataset02_10}
\end{table}

\subsection{ Results on Stability of Features}

In this section, we experiment on the stability in the feature selection process so that the method can retrieve consistent features across random subsamplings of the dataset. 
The stability of selected features is a critical aspect in feature selection tasks. This study helps us to measure how consistently the same features are selected across different iterations, datasets, or models. A stable feature selection method indicates robustness and reliability. 

In the bottom row of Figure \ref{fig:dataset1}, we show the stability index as a function of the percentage of the feature-component selected for Dataset 1. Notice that LASSO and Relief selections have competitive performance in terms of the Kuncheva index, whereas LASSO consistently outperforms for the Jaccard index.  
In the bottom row of Figure \ref{fig:dataset2}, we show the stability index as a function of the percentage of the feature-component selected for Dataset 2. We found that LASSO consistently outperforms for indexes.  

In addition to the graphical analysis, we present the mean stability ratings for each feature selection method based on 100 cross-validation subsamplings.  On Dataset 1, LASSO had the most stability, with an average Kuncheva Index of around 0.69 and a Jaccard Similarity of 0.62.  Relief achieved ratings of 0.64 (Kuncheva) and 0.59 (Jaccard), whereas ANOVA recorded scores of 0.61 and 0.55, respectively.  In Dataset 2, the trend remained consistent, with LASSO producing the most stable feature subsets, with an average Kuncheva Index of 0.74 and a Jaccard Similarity of 0.69.  Relief and ANOVA earned Kuncheva scores of 0.70 and 0.68, as well as Jaccard values of 0.65 and 0.63, respectively.  These findings validate that LASSO not only promotes classification efficacy but also improves feature stability, hence affirming its appropriateness for reproducible biomarker identification.

\section{Identification of Discriminatory Brain Regions}
\label{sec:explain}

In this work, we analyze how the varying brain regions influence the classification of brain disorders. We delve into a targeted analysis of the regions of interest (ROI) within the brain. By examining the degree of relevance attributed to specific ROIs, we identified what we refer to as the 'affected core'. This core consists of ROIs that consistently exhibit significant relevance, suggesting their critical role in the underlying condition being studied.

Following feature selection, we identified the most frequently selected brain regions across folds and random seeds. Selected features, representing vectorized upper-triangular entries of the connectivity matrix, were mapped back to ROI pairs using the inverse of the standard upper-triangle index mapping. For each ROI, we computed its selection frequency as the number of times it appeared in any selected feature pair.
From the top 10 ROIs selected by the feature selection methods in both hemispheres, we observed several regions with potential relevance to schizophrenia for both data sets in our experiments.

For the identification of 'affected core' regions, we apply an indicator function-based rule. In our experiment, we have 5-fold cross validation in each trial, consisting of 7 different values of percentage of selected feature-component: $[0.5, 1, 2, 5, 10, 25, 50]$. We repeated the experiment over 10 trials. Now, during feature selection at each trial and each validation, we update the frequency counter of regions $r$-th and $t$-th if the edge $(r, t)$ is selected in the LASSO/Relief/ANOVA step. With initialization of $\delta_r = 0$ and $\delta_t = 0$, we update the frequency counter as follows:
\begin{equation}
\begin{split}
     \delta_r = \delta_r + 1  \\
     \delta_t = \delta_t + 1    
\end{split}
\Bigg\} \ \text{if edge }  \ (r,t) \ \text{is selected}.   
 \end{equation}

In Fig. \ref{fig:affected_ds1}, we display the top 10 selected brain regions from the left hemisphere and the right hemisphere from Dataset 1 using LASSO, ANOVA, and Relief selections. 
Notice that for LASSO selection, the common (between left and right hemispheres) affected core regions are: (i) superior-temporal, (ii) middle-temporal, (iii) posterior-cingulate, (iv) pre-central, and (v) post-central.
%
For ANOVA selection, the common 'affected core' brain regions are: post-central, isthmus-cingulate, and fusiform. For Relief selection, the common 'affected core' brain regions are: transverse-temporal, para-central, putamen, amygdala, and posterior-cingulate. Now considering higher classification performance and stability index of LASSO selection, we infer that the top 5 'affected core' brain regions are: superior-temporal, middle-temporal, posterior-cingulate, pre-central, and post-central.


\section{Conclusion}
\label{sec:conc}

In this research, we studied the effectiveness of feature selection algorithms for functional connectivity-based disease categorization in the brain.
We tested LASSO, Relief, and ANOVA on vectorized fMRI connectomes for classification and feature stability throughout cross-validation folds. We showed that LASSO consistently has the best accuracy, reproducibility trade-off using stability criteria like the \textit{Kuncheva index} and \textit{Jaccard index}. We found that mapping selected attributes to brain regions of interest enhanced both interpretability and performance evaluation. This allowed us to discover consistently selected and analytically acceptable brain regions that may contribute to disorder-related neural patterns. These results enable us to make connectome-based biomarker detection models clearer and dependable. Our results are validated using two separate fMRI datasets, comprising a total of 226 participants, indicating adaptability in both classification accuracy and feature stability. In future research, we intend to investigate the integration of multimodal data sources, including EEG and MEG, to capture and enhance temporal dynamics. Furthermore, broadening this approach to include multiclass classification scenarios and other clinical populations (e.g., ADHD, bipolar disorder, early cognitive impairment) could further augment its application.  Ultimately, integrating neurobiological priors or attention mechanisms into the feature selection process could produce more interpretable and robust models for therapeutic application.

\section*{Acknowledgment}
The authors are grateful to the researchers who
made the open-access neuroimaging datasets \cite{GUTIERREZGOMEZ2020102316}, \cite{poldrack2016phenome, Sunil2024} that were used in this work.  The University of Lausanne, Switzerland, shared Dataset 1, and UCLA made Dataset 2 open to everyone.  These datasets made it possible to do this study without the collection of clinical data.

\bibliographystyle{IEEEtran}
\bibliography{ref}

\end{document}